# Bose-Einstein condensation model for high-temperature superconductivity

A. Rosencwaig*

*Department of Physics, University of Toronto, Toronto, Ontario, Canada M5S 1A7*

I propose that a dopant charge singlet bonding state may arise from the hybridization of molecular orbitals in a cluster containing 13 Cu atoms in the $CuO_2$ plane of the superconducting cuprates. This singlet state forms a preformed pair with low binding energy that is spatially bounded and weakly interacting, and that can undergo Bose-Einstein condensation. I show that this model is able to account, in a quantitative and natural way, for many of the thermodynamic and electronic characteristics of the superconducting cuprates, including many of the key experimental angle-resolved photoemission spectroscopy, muon spin-relaxation, and microwave results on the temperature and doping dependencies of both the superfluid density and the pairing strengths (superconducting gap, leading-edge midpoint, and pseudogap) in these high-temperature superconductors.

## I. THE $(Cu)_{13}$ CLUSTER AND THE SINGLET STATE

Since the discovery of high-temperature cuprate superconductors by Bednorz and Muller in 1986,[1] there has been a great deal of experimental and theoretical work to explore and explain the phenomenon of high-temperature superconductivity. The wealth of experimental data obtained over the past 16 years has provided a comprehensive catalog of the properties of this new class of superconductors. It has, however, been particularly challenging to develop a theory of high-temperature superconductivity that can account for all of the various experimental observations, although many approaches have been tried. In spite of considerable progress, a complete understanding of high-temperature superconductivity still remains elusive.

Superconductivity in the cuprates is intimately related to the presence of dopant charge (electrons or holes) in the $CuO_2$ planes or layers. A Cu atom in a $CuO_2$ plane is covalently bonded to four O atoms in the plane, and the highest filled molecular orbital of this $(CuO_4)^{-6}$ cluster is the antibonding $\sigma^*_{x^2-y^2} = d_{x^2-y^2} \mp p\sigma_x \pm p\sigma_y$ orbital. In the undoped tetragonal cuprates, the crystal field and strong on-site Coulomb correlation splits the half filled $\sigma^*_{x^2-y^2}$ band into an empty Hubbard conduction band primarily of Cu $3d$ character and a filled valence band primarily of O $2p$ character separated by a charge-transfer gap of 1–2 eV. In Fig. 1, we depict a central Cu atom, $A$, covalently bonded through O atoms to its four nearest Cu neighbors, $B_1$, $B_2$, $B_3$, and $B_4$ all at a distance $a$, the lattice constant in the plane. We also show the second-nearest Cu neighbors, $C_1$, $C_2$, $C_3$, and $C_4$ that are all at a distance $\sqrt{2}a$. We note that the phases or symmetries of the $\sigma^*_{x^2-y^2}$ molecular orbitals at the $B$ sites are all the same but opposite to that at site $A$, while the $C$ site orbitals have the same symmetry as site $A$. In Fig. 2, we expand on the number of Cu sites displayed but rather than draw all of the antibonding orbitals we simply indicate the symmetry of the orbitals at various Cu sites with a $(+)$ or $(-)$.

There have been a number of molecular-orbital and band-structure treatments of dopant charges in the $CuO_2$ planar structure. These include the single-band Hubbard model where only charges in the Cu $d$ orbitals are considered,[2] and the two-band $t$-$J$ models that include the O $2p$ orbitals as well.[3–5] The models that appear to best agree with angle-resolved photoemission spectroscopy (ARPES) data are $t$-$t'$-$t''$-$J$ models where long-range charge hopping, up to third-nearest neighbors, for Cu-O and O-O are considered.[6,7] We propose a similar $t$-$t'$-$t''$-$J$ treatment where we consider direct long-range Cu-Cu hopping between Cu atoms. In determining the site-hopping matrix element $t$ for charge hopping between a central Cu site ($A$) and nearest-neighbor ($B$), next-nearest neighbor ($C$), and third-nearest neighbor ($D$) Cu sites, we assume that distance, orbital orientation, and orbital symmetry all play important roles. Thus we see from Fig. 2 that $A$-$B$ interactions have favorable distance $a$, favorable orbital orientation, but unfavorable symmetry; $A$-$C$ interactions have less favorable distance $\sqrt{2}a$, highly unfavorable orbital orientation (the sites being on a diagonal while the orbitals are aligned along $x$ and $y$), but favorable symmetry; while the third-nearest neighbor $A$-$D$ interactions have unfavorable distance $2a$, but favorable orientation, and favorable symmetry. Interactions with Cu sites beyond the third-nearest $D$ neighbors are all much reduced because of unfavorable orientation and symmetry as well as increasing distance. Combining the effects of distance, orbital orientation, and symmetry it is reasonable to assume that $A$-$B$, $A$-$C$, and $A$-$D$ exchange integrals are all comparable but considerably greater than exchange integrals beyond the $D$ neighbors. Thus the primary dopant charge interaction cluster consists of the central Cu, $A$, the four nearest Cu $B$ neighbors, the four second-nearest Cu $C$ neighbors, and the four third-nearest Cu $D$ neighbors. This then constitutes a fundamental interaction cluster of 13 Cu atoms and 26 O atoms, which we designate as the $(Cu)_{13}$ cluster.

In this $t$-$t'$-$t''$-$J$ model, dopant interactions between the 13 Cu sites in the $(Cu)_{13}$ cluster can result in a hybridization of the 13 dopant antibonding $\sigma^*_{x^2-y^2}$ molecular orbitals into 13 dopant cluster orbitals, which can be considered to constitute a miniature impurity band located near the top of the undoped valence band for $p$ doping and near the bottom of the undoped conduction band for $n$ doping.[8] By hybridizing the $\sigma^*_{x^2-y^2}$ orbitals, we include O $2p$ as well as Cu $3d$ orbitals in the charge hopping process. Antiferromagnetic exchange interactions between the dopant charge and the valence electron on the Cu sites will make singly occupied and triplet cluster states more energetic than singlet states because of spin flipping as the charge moves from site to site within the cluster. The singlet states, which need not undergo spin flipping, are thus the lowest-energy states. These singlet states are formed from two dopant charges and thus are not Zhang-Rice singlets[3] that are formed from one dopant charge with opposite spin to a local Cu spin. The complete energy-level diagram for the dopant cluster orbitals must of course be developed from a full molecular-orbital treatment. However, we can, with no loss of generality, assume for simplicity that the 13 singlet cluster orbitals are spaced evenly apart by an energy $\delta$, as depicted by the $n_n$ states in Fig. 3. The six lower states are bonding states, the center is a nonbonding state, and the upper six states are antibonding states. We propose that the lowest-energy singlet $(\psi_0)^2$ state will have a primarily $d_{x^2-y^2}$ symmetry since all the constituent $\sigma^*_{x^2-y^2}$ molecular orbitals have this symmetry component and the $p\sigma_x$ and $p\sigma_y$ orbitals will average out. The symmetry of the energy gap $\delta$ will also be $d_{x^2-y^2}$. All six of the bonding singlet states are spatially bounded and thus have a fair amount of phase coherence. However, to qualify as bosonic quasiparticles for purposes of superconductivity, a singlet state must retain its coherence during transport, i.e., during the hopping process between clusters. We will assume that only the ground $(\psi_0)^2$ singlet state retains enough phase coherence during transport to qualify as a bosonic quasiparticle, or preformed pair, that can participate in superconductivity.

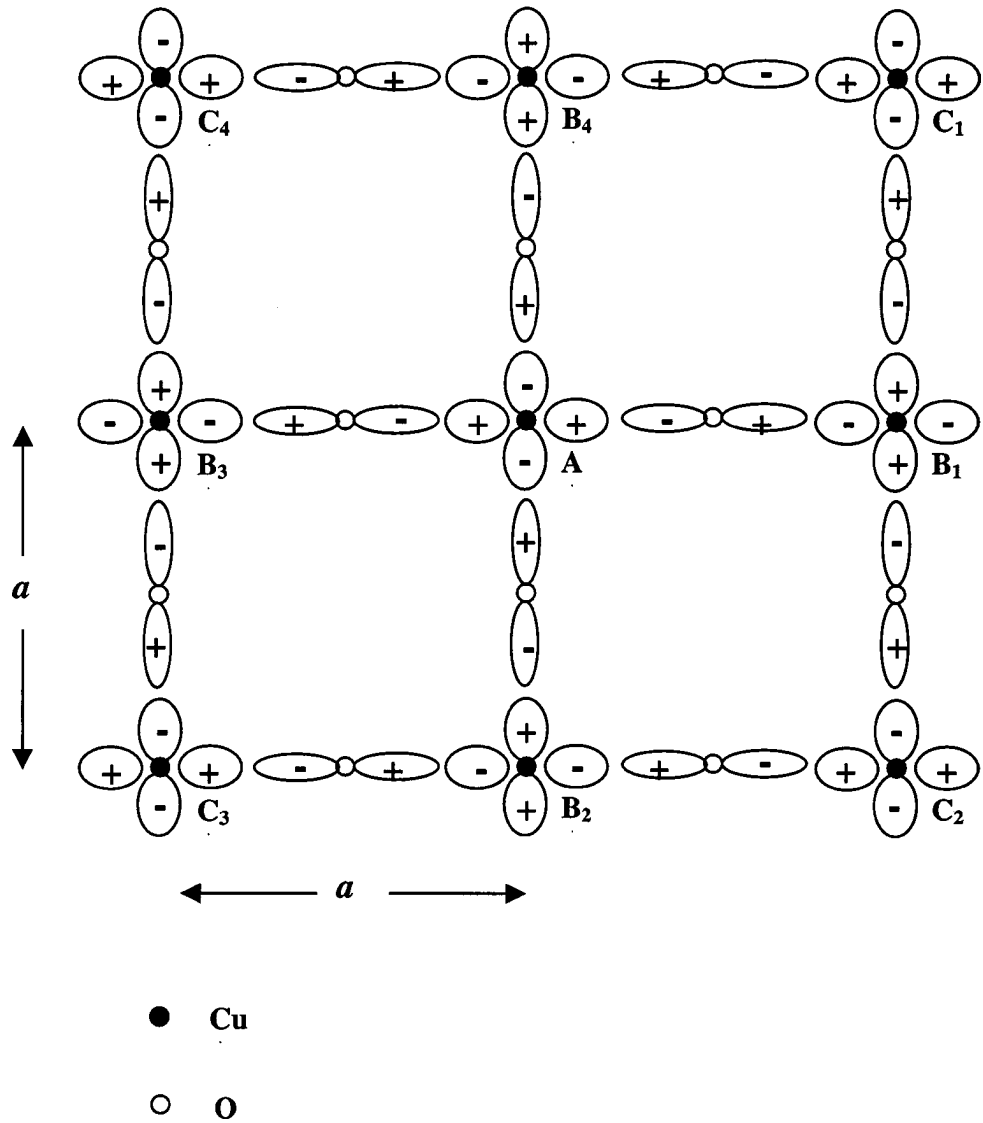

FIG. 1. Schematic representation of the square-planar $CuO_2$ layer showing a central Cu atom $A$, the four $B$ nearest neighbors, and the four $C$ next-nearest neighbors, together with the in-between O atoms. The Cu $d_{x^2-y^2}$ and O $2p\sigma$ atomic orbitals making up the antibonding $\sigma^*_{x^2-y^2}$ molecular orbitals are shown, with the wave-function phase indicated.

This qualitative treatment to obtain a bonding singlet state differs from previous treatments in several respects. First, the interaction cluster is much larger, consisting of 13 Cu atoms, and we assume direct Cu-Cu charge hopping up to third-nearest neighbors. Secondly, we propose that when the electrostatic potential from the dopant ions is included in the crystal Hamiltonian, the $(\psi_0)^2$ singlet state will be spatially bounded to dimensions comparable to the size of the (Cu)$_{13}$ cluster. Furthermore, the singlet state will interact only weakly with other singlet states because the potential from the dopant ions will effectively screen the local $(\psi_0)^2$ charge pair from the Coulomb repulsion of charge pairs from other clusters. The relatively weak interaction between the singlet charge pairs will consist of some residual Coulomb repulsion and a site-hopping interaction when the outer Cu atoms from two clusters are within a distance $l$ apart where $a \le l \le 2a$. For simplicity we will assume that $l \approx 1.5a$. This then puts the interaction distance between two singlet $(\psi_0)^2$ states at $d = 4a + l \approx 5.5a$. As we add more clusters and assemble the lattice, the singlet state is delocalized a little and its wave function reaches a diameter $d$ because of the site-hopping interactions between adjacent clusters. However, it still remains spatially bounded and does not broaden out into a wide band.

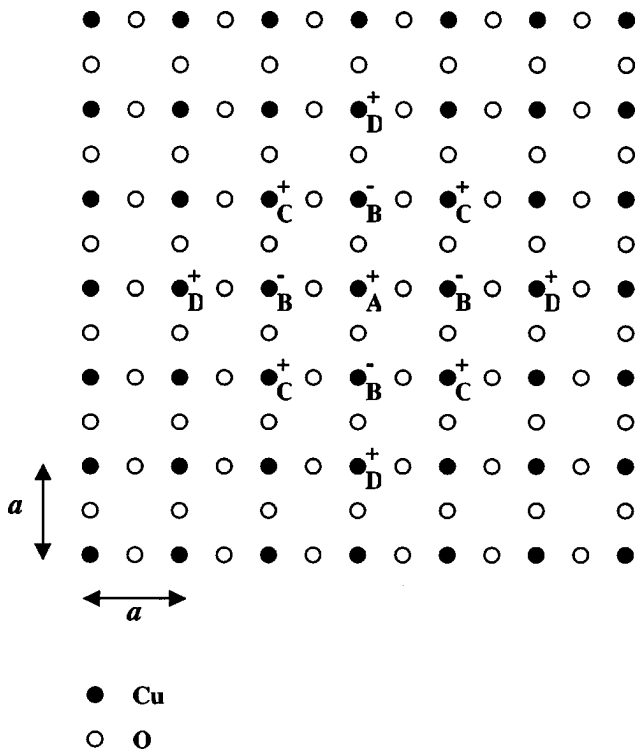

FIG. 2. Schematic representation of a larger section of the CuO$_2$ plane. The relative wave-function phases or symmetries of the $\sigma^*_{x^2-y^2}$ orbitals for various Cu sites are indicated by ($+$) and ($-$).

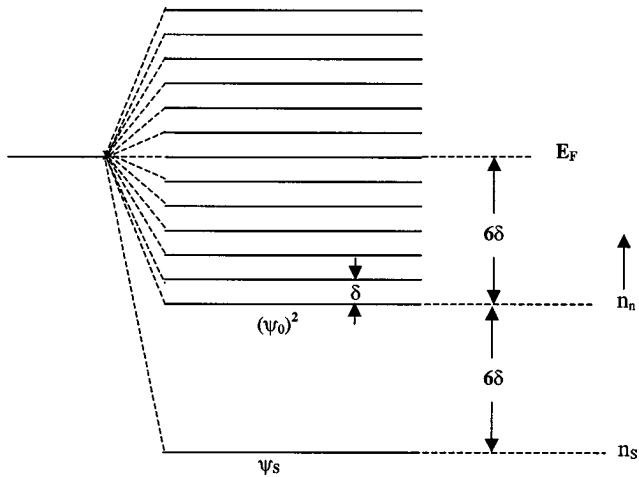

FIG. 3. Hypothetical simplified energy-level diagram for the 13 dopant cluster orbitals formed from the hybridization of the 13 $\sigma^*_{x^2-y^2}$ molecular orbitals in the (Cu)$_{13}$ cluster. The lowest bonding singlet orbital is designated by $(\psi_0)^2$ and its single-particle energy is at $-6\delta$ relative to the Fermi level $E_F$, which is at the center nonbonding orbital of the normal states $n_n$. The superconducting coherent quasiparticle state $\psi_s$ is located at $-12\delta$.

The singlet $(\psi_0)^2$ state is a hole or electron pair and represents a charged bosonic quasiparticle, or preformed pair, with a diameter of $\approx 5.5a$ or $\approx 20$ Å since $a$ in the cuprates is typically $\approx 3.8$–$3.9$ Å. Because of the screening potential of the dopant ions, these charge pairs can move through the lattice as preformed pairs that are weakly interacting. In addition, as we shall see later, these charge pairs have densities that are temperature dependent, and have low binding energies and thus are not hard-core bosons. These characteristics of the $(\psi_0)^2$ charge pairs may possibly account for the non-Fermi-liquid behavior of the normal state of the cuprates. Of most importance, however, is that because of the spatially bound nature of the $(\psi_0)^2$ state and the weak interactions between these bosonic quasiparticles, they can, under the right conditions, experience Bose-Einstein condensation and thus generate a superconducting state. This is not the first time that a Bose-Einstein condensation of preformed pairs has been suggested as the underlying mechanism at work in the high-temperature superconductors. Uemura *et al.* have proposed this mechanism to account for the celebrated relationship between $T_c$ and the muon spin-relaxation rate at $T \to 0$.[9,10] We return to the Uemura relationship later in the paper.

## II. BOSE-EINSTEIN CONDENSATION

The quasiparticles, represented by the $(\psi_0)^2$ singlet states, are spatially bound weakly interacting bosons with an interaction distance $d \approx 5.5a$. This situation lends itself to a Bose-Einstein condensation (BEC) condition.[11] We will assume that the three-dimensional BEC condition for a weakly interacting system of bosons is applicable. Thus,

$$n_b \lambda^3 = 2.612, \quad (1)$$

where $n_b$ is the boson density, $\lambda$ is the wavelength (as a diameter) of the bosonic quasiparticle, and Eq. (1) simply states that BEC occurs when the average distance between the bosons becomes somewhat less than the wavelength or interaction distance. In conventional BEC the boson density is independent of temperature $T$, while $\lambda$ is the thermal wavelength and is dependent on $T$. Here, it is the opposite.

The wavelength is the interaction distance $d$ and is essentially independent of $T$, but the boson density is dependent on $T$, since it depends on the occupation probability of the singlet $(\psi_0)^2$ state. This is a major element of the $(Cu)_{13}$-BEC model. The boson density $n_b$ is related to the cluster density $N_c$ by $n_b = N_c n_{bc}$, where $n_{bc}$ is the number of bosons per cluster. If we define the primitive tetragonal cell as having a volume $(a \times a \times c)$ Å$^3$, where $a$ and $c$ are the tetragonal lattice constants of the primitive unit cell, and since each primitive unit cell contains only one formula unit and only one Cu atom per $CuO_2$ plane, then $N_c = 1/(13a^2c)$. If $q$ is the average dopant charge per Cu atom in a layer of the cluster, and if there are $n$ $CuO_2$ layers per cluster, then the total charge in the cluster is $m = 13nq$. The number of bosons in the cluster is then $n_{bc} = \frac{1}{2}(13nq)P(\delta,T)$, where $P(\delta,T)$ is the probability that a charge is in the $\psi_0$ state at temperature $T$ and $\delta$ is the energy gap separating the $\psi_0$ state from the next state. $P(\delta,T)$ is then given by

$$P(\delta,T) = 1/[1 + e^{-\delta/kT} + e^{-2\delta/kT} + \ldots + e^{-12\delta/kT}]. \quad (2)$$

Substituting $\lambda = 5.5a$, we then find for the basic condition for Bose-Einstein condensation in the cuprates

$$13qP(\delta,T) = 0.408(c/a)/n. \quad (3)$$

Since $13qP(\delta,T)$ represents the number of charges in the $\psi_0$ state per $CuO_2$ layer in a cluster, and since there is only one $\psi_0$ state per $CuO_2$ layer in the cluster, a second condition is

$$13qP(\delta,T) \leq 2. \quad (4)$$

We have found that, in spite of the stringent BEC conditions, all of the cuprates readily satisfy Eqs. (3) and (4) for all experimental values of $a$, $c$, $n$, and $q$, and do so at very reasonable values of $P(\delta,T)$. From Eq. (3) we can make the following general comments. For a constant $\delta$, a decreasing $P(\delta,T)$ results in an increasing $T = T_c$. Thus, we expect $T_c$ to increase with decreasing $(c/a)$ and increasing $n$. In addition, $T_c$ will increase with increasing charge per Cu, $q$. However this will not continue indefinitely, since as we discuss below, $\delta$ is itself a decreasing function of $q$.

In this $(Cu)_{13}$-BEC model, the superconducting charge pair is simply the $(\psi_0)^2$ preformed pair condensed into the superconducting state. The symmetry of the superconducting state will thus be the same as that of the preformed $(\psi_0)^2$ state, i.e., $d_{x^2-y^2}$. This is in agreement with experiment. Furthermore, since the interaction distance or wavelength $\lambda$ of the preformed pair is independent of temperature, the extent (as a radius) of the wave function of the superconducting pair at $T=0$, and hence the correlation length at $T=0$, $\xi_0$, is simply $= \frac{1}{2}\lambda = 2.75a \approx 10$ Å, also in agreement with experiment. Therefore both the observed symmetry and the correlation length of the superconducting pairs in the cuprates follow naturally from the $(Cu)_{13}$-BEC model.

### III. $T_c$ vs $q$ CURVES

At low values of $q$, Eq. (4) is always satisfied, and Eq. (3) will give the threshold dopant charge concentration per Cu atom at which BEC and thus superconductivity can occur. At the threshold concentration, BEC occurs at very low temperatures, and thus $P(\delta,T) \approx 1$, for any $\delta$. In addition, as long as the dopant charge is mobile, we can have two-hole clusters ($m=2$) even for low average values of $q$. This is possible because when two one-hole clusters come within an interaction distance of each other, there will be a preferential formation of one two-hole cluster because of the lower energy of the singlet state. Since most cuprates have mobile dopant charges at low $q$, we find for the threshold value $q_0 = 0.408(c/a)/13n$. The situation for $n$-doped $Nd_{2-x}Ce_xCuO_4$ (NCCO) is quite different in that the dopant electron is not mobile at very low $q$ concentrations since NCCO is an insulator until $q \approx 0.14$.[12] Thus we expect only one-electron clusters to form up to $q = \frac{1}{13}$ (0.077). We thus modify Eq. (3) to take this into account and obtain for NCCO $2(q_0 - 0.077) = 0.408 (c/a)/13n$.

We list in Table I the calculated and experimental (measured or estimated) values of the threshold dopant value $q_0$ for several representative cuprates. Experimental values of $q_0$ appear to exist only for $La_{2-x}Sr_xCuO$ (LSCO) (Ref. 14) and $YBa_2Cu_3O_{7-y}$ (Y123),[15] the others being estimates of $\approx 0.05$.[15] We find excellent agreement between theory and experiment for LSCO and Y123, and good to excellent agreement with the estimated $q_0$'s for the other cuprates. There is no experimental value or estimate of the $q_0$ for NCCO since this compound becomes an abrupt superconductor at the insulator-metal transition at $q \approx 0.14$. The theoretical value of 0.10 represents the effective threshold value. We note that the calculated threshold levels $q_0$ are determined solely by readily measurable quantities $a$, $c$, and $n$, and are totally independent of the variable $\delta$.

To obtain the full dependence of the transition temperature $T_c$ on dopant charge per in-plane Cu, $q$, we need to obtain values for the energy gap $\delta$ and its dependence on $q$. First let us determine what the model and the physics of the problem say about this dependence. The energy gap $\delta$ is affected by local crystal fields and by the Coulomb repulsion of local charges, and since both the dopant ion concentration (local fields) and the local charge are proportional to $q$, $\delta$ must itself be a function of $q$. Due to strong electron correlations and strong on-site Coulomb repulsion in the $CuO_2$ planes, it is reasonable to assume that the probability for forming a stable bound preformed pair in the $(Cu)_{13}$ cluster will decrease with increasing $q$. In the model this is equivalent to assuming that $\delta$ will decrease with $q$. It is also reasonable to assume that, at some higher value of $q$, the number of charges in the cluster is large enough that the strong Coulomb repulsion prevents the formation of stable bound preformed pairs. This is equivalent in the model to setting $\delta \to 0$ since then the cluster energy manifold will collapse and we will be left with only nonbonding states. Thus, without resorting to any fitting procedure, the physics and the model tell us that $\delta$ must decrease with $q$ and that, at some upper value of $q$, $\delta \to 0$. From an analysis of experimental $T_c$ vs $q$ curves, we have found that $\delta$ appears to simply decrease linearly with $q$ throughout the superconducting dopant range. Thus it is possible to calculate $\delta(q)$ from only two data points on a $T_c$ vs $q$ experimental curve.

TABLE I. Threshold doping for various cuprates. The lattice constants $a$ and $c$ are for the primitive unit cell, which contains only one formula unit and one Cu atom per layer, with $n$ $CuO_2$ layers in the cell. The experimental threshold doping concentrations, $q_0$(exp), are from Refs. 14 and 15, and the theoretical threshold doping concentrations, $q_0$(th), are calculated from the $(Cu)_{13}$-BEC model.

| Cuprate | $a$ (Å) | $c$ (Å) | $n$ | $q_0$(exp) | $q_0$(th) |
|---|---|---|---|---|---|
| $La_{2-x}Sr_xCuO_4$ (LSCO) | 3.78 | 6.60 | 1 | 0.056 | 0.055 |
| $Nd_{2-x}Ce_xCuO_4$ (NCCO) | 3.94 | 6.05 | 1 | | $\approx 0.10$ |
| $YBa_2Cu_3O_{7-y}$ (Y123) | 3.85 | 11.65 | 2 | 0.05 | 0.047 |
| $Bi_2Sr_2CaCu_2O_{8-y}$ (Bi2212) | 3.9 | 15.4 | 2 | $\approx 0.05$ | 0.062 |
| $Bi_2Sr_2Ca_2Cu_3O_{10-y}$ (Bi2223) | 3.9 | 18.6 | 3 | $\approx 0.05$ | 0.050 |
| $HgBa_2CaCu_2O_{7-y}$ (Hg1212) | 3.9 | 12.7 | 2 | $\approx 0.05$ | 0.051 |
| $HgBa_2Ca_2Cu_3O_{9-y}$ (Hg1223) | 3.9 | 15.9 | 3 | $\approx 0.05$ | 0.043 |

TABLE II. Experimental and theoretical parameters for various cuprates. The maximum $T_c^m$ is at the optimal doping $q_m$. The theoretical $T_c^m$, $q_m$, and energy gap $\delta_m$ are derived from the $(Cu)_{13}$-BEC model. The experimental values for $\Delta_m(\exp)$ are from Ref. 25 and the estimated values are derived from Ref. 25; the theoretical $\Delta_m(\text{th}) = 12\delta_m$.

| Cuprate | $T_c^m(\exp)$ (K) | $T_c^m(\text{th})$ (K) | $q_m(\exp)$ | $q_m(\text{th})$ | $\delta_m$ (meV) | $\Delta_m(\exp)$ (meV) | $\Delta_m(\text{th}) = 12\delta_m$ (meV) |
|---|---|---|---|---|---|---|---|
| LSCO | 36 | 36 | ≈0.15 | 0.154 | 1.35 | ≈20 (estimated) | 16.2 |
| NCCO | 22 | 22 | 0.14–0.15 | ≈0.14 | 0.80 | ≈12 (estimated) | 9.6 |
| Y123 | 92 | 93 | ≈0.16 | ≈0.15 | 2.91 | ≈42 | 34.9 |
| Bi2212 | 95 | 96 | ≈0.16 | ≈0.15 | 4.22 | ≈40 | 50.6 |
| Bi2223 | 110 | 111 | ≈0.16 | ≈0.15 | 3.25 | ≈45 | 39.0 |
| Hg1212 | 127 | 128 | ≈0.16 | ≈0.15 | 4.39 | ≈52 | 52.7 |
| Hg1223 | 133 | 135 | ≈0.16 | ≈0.15 | 3.65 | ≈54 (estimated) | 43.8 |

For the cuprates, $T_c$ decreases and the superconductivity completely disappears in the overdoped region beyond which the material then behaves as a normal metal. This behavior is readily explained by the $(Cu)_{13}$-BEC model with the assumption that, at some value of $q$, $\delta \to 0$ because of strong Coulomb repulsion. In the model, the disappearance of superconductivity in the overdoped region occurs when $\delta \to 0$, since at this point, all 13 cluster states become degenerate and nonbonding with the result that there can be no bound bosonic quasiparticles at any temperature. The system can therefore no longer support superconductivity. In addition, since there are no longer any preformed pairs, the system will now behave as a conventional Fermi-liquid metal. Thus we can clearly identify the point on the $T_c$ vs $q$ curve in the overdoped region where $T_c \to 0$ as the one where $\delta \to 0$ as well. This then is a singular point on the $T_c$ vs $q$ curve and we will use it as one of the fitting points to determine $\delta(q)$.

For the other point, we generally use $q = \frac{2}{13}$ ($m = 2$), since at this doping all of the clusters have exactly two charges.

We list in Table II the experimental and theoretical values for the maximum $T_c$, $T_c^m$, the dopant charge $q$ at $T_c^m$, $q_m$, and the calculated values of the energy gap $\delta$ at $T_c^m$, $\delta_m$, for the representative cuprates. In Fig. 4 we show the theoretical $T_c$ vs $q$ curves for both LSCO and NCCO together with experimental data.[12–14] We see very good agreement in both cases. Of particular interest is that in LSCO the change of $T_c$ with $q$ is very abrupt near the threshold doping level. However, this behavior appears to be supported by recent data from Fujita *et al.*[14] We find that the theoretical $q_m$ for LSCO is ≈0.15 ($\approx \frac{2}{13}$), in agreement with the experimental value of 0.15. For NCCO, the theoretical curve, which peaks at ≈0.14, clearly shows why this material becomes abruptly superconducting at the insulator-metal transition. When the dopant charge becomes mobile at $q \approx 0.14$, the system suddenly finds itself with a boson density well in excess of the threshold value needed for BEC, since that threshold density is reached, according to the theory, at $q = 0.10$, and so the system abruptly becomes superconducting with a $T_c$ appro-priate for the higher boson density, as indicated by the theoretical curve.

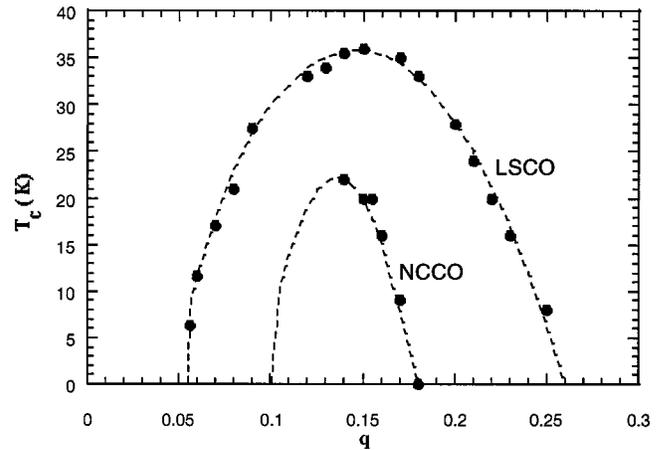

FIG. 4. The transition temperature $T_c$ vs the dopant charge per in-plane Cu $q$ for $La_{2-x}Sr_xCuO_4$ (LSCO) and $Nd_{2-x}Ce_xCuO_4$ (NCCO). The experimental data are from Refs. 12–14. The *n*-doped cuprate NCCO undergoes an insulator-metal transition at $q \approx 0.14$, at which point it becomes abruptly superconducting.

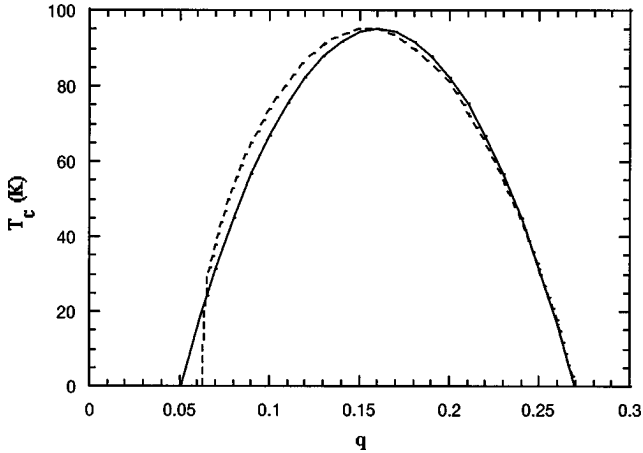

FIG. 5. The transition temperature $T_c$ vs the dopant charge per in-plane Cu $q$ for $Bi_2Sr_2CaCu_2O_{8-y}$ (Bi2212). The curve (solid line) is derived from the empirical formula $T_c/T_c^m = 1 - 82.6(q-0.16)^2$ (Ref. 16) with $T_c^m = 95$ K. The theoretical curve is indicated by a dashed line.

TABLE III. A sensitivity analysis for the calculated $q_m$, $T_c^m$, and $\delta_m$ as a function of fitting points $(q_1, q_2)$ used to obtain the linear $\delta$ vs $q$ curve.

| LSCO | | | | |
|---|---|---|---|---|
| $(q_1, q_2)$ | (0.10, 0.26) | $(\frac{2}{13}, 0.26)$ | (0.20, 0.26) | (0.10, 0.20) |
| $q_m$ | 0.148 | 0.150 | 0.150 | 0.149 |
| $T_c^m$ (K) | 36 | 36 | 35.5 | 36 |
| $\delta_m$ (meV) | 1.44 | 1.41 | 1.41 | 1.42 |
| Bi2212 | | | | |
| $(q_1, q_2)$ | (0.10, 0.27) | $(\frac{2}{13}, 0.27)$ | (0.20, 0.27) | (0.10, 0.20) |
| $q_m$ | 0.153 | 0.154 | 0.154 | 0.153 |
| $T_c^m$ (K) | 90 | 95 | 96 | 92 |
| $\delta_m$ (meV) | 4.00 | 4.20 | 4.26 | 4.08 |

Most p-doped cuprates are assumed to have phase diagrams, and thus $T_c$ vs $q$ curves, similar to LSCO.[15] Thus a generic empirical relationship[16] is generally assumed for most cuprates, with $T_c/T_c^m = 1 - 82.6(q-0.16)^2$. In Fig. 5 we show the $T_c$ vs $q$ curve obtained for Bi2212 from the $(Cu)_{13}$-BEC model together with the curve derived from the empirical expression. Although there is, in general, good agreement between the two curves there are some interesting points of comparison. First, the theoretical curve has a threshold $q_0$ for Bi2212 of $\approx 0.06$ whereas the empirical curve puts it at 0.05. Secondly, while the empirical curve is a symmetric parabola, the theoretical curve is asymmetric, rising faster in the underdoped region and falling at the same rate as the empirical curve in the overdoped region. As in LSCO, the increase in $T_c$ with $q$ is especially abrupt near the doping threshold level. Of particular interest is the fact that the theoretical curve for Bi2212, and in fact for most cuprates, tends to peak at $q \approx 0.154$ $(\frac{2}{13})$, which is quite close to the empirical value of 0.16. This behavior is to be expected from the $(Cu)_{13}$-BEC model since a $q$ of $\frac{2}{13}$ represents the situation in which all of the $(Cu)_{13}$ clusters have exactly two charges per $CuO_2$ layer and thus one preformed pair and one superconducting pair at $T \to 0$. Since there can be only one preformed pair per layer in a cluster, this value of $q$ represents the maximum possible superfluid density or stiffness at $T \to 0$. As $q$ increases beyond $\frac{2}{13}$, there can be no further increase in superfluid density at $T \to 0$, but there is a continuing decrease in $\delta$ and thus in the superconducting pairing strength. Thus while $T_c$ can increase with increasing $q$ below $q = \frac{2}{13}$, it must begin to decrease for $q > \frac{2}{13}$. The general observation that superconducting cuprates tend to have bell-shaped doping curves with maximum transition temperatures in the region of $q = 0.15 - 0.16$ is thus a natural consequence of the $(Cu)_{13}$-BEC model.

An obvious concern is the sensitivity of the theoretical $T_c$ vs $q$ curve to the specific fitting protocol used to establish the $\delta$ vs $q$ curve and, in particular, whether the values of the key parameters $q_m$, $\delta_m$, and $T_c^m$ are predetermined by our selection of $q = \frac{2}{13}$ as one of the two fitting points. Table III presents a sensitivity analysis for both LSCO and Bi2212 using four different fitting protocols for each material. For LSCO we have obtained the $\delta$ vs $q$ curve using for the two fitting points $(q_1, q_2)$, (0.10, 0.26), (2/13, 0.26), (0.20, 0.26), and (0.10, 0.20). For Bi2212 we have used as fitting points (0.10, 0.27), (2/13, 0.27), (0.20, 0.27), and (0.10, 0.20). We see from Table III that $q_m$ is especially robust for both materials, varying by less that 2% for the different fitting protocols, thus verifying that the $T_c$ vs $q$ curve is indeed bell shaped with a peak at $q_m \approx \frac{2}{13}$. The $T_c^m$ and $\delta_m$ are a bit more sensitive to the fitting protocol, although here again the variation in $T_c^m$ and $\delta_m$ is <5% in LSCO and <6% in Bi2212. We can thus conclude that the values of $q_m$, $T_c^m$, and $\delta_m$ recorded in Table II are indeed quite robust and the apparent good fits to the data are not an artifact of the $\delta$ vs $q$ fitting process.

One of the key features of the superconducting cuprates is the fact that within a homologous series of cuprates, such as the 1-Tl, 2-Tl, 2-Bi, and 1-Hg series of compounds, $T_c^m$ tends to increase with the number $n$ of $CuO_2$ layers but then begins to saturate, or in some cases decrease, with increasing $n$. This behavior is readily explained by the $(Cu)_{13}$-BEC model. The lattice constant $c$ in the primitive cell changes as $n$ increases within a given homologous series as $c_n = c_1 + (n-1)t$, where $t$ is the distance between adjacent $CuO_2$ planes. Substituting this expression for $c$ in Eq. (3), we have at optimal doping

$$13 q_m P(\delta_m, T_c^m) = 0.408\{[c_1 + (n-1)t]/a\}/n. \quad (5)$$

We can see that the right-hand side of Eq. (5) will decrease with increasing $n$. If both $q_m$ and $\delta_m$ remain constant, then this decrease of the right-hand side will result in a decrease in $P(\delta_m, T_c^m)$ and thus in an increase in $T_c^m$. However as $n$ continues to increase, the decrease in the right-hand side saturates toward a constant value of $0.408(t/a)$, and thus $T_c^m$ will saturate toward a constant value as well. If, however, $q_m$ or $\delta_m$ decreases with increasing $n$, then $T_c^m$ will reach a maximum level and then begin to decrease as $n$ continues to increase. The dopant charge $q_m$, averaged over the layers, can decrease as $n$ increases because the inner layers may get less charge than the outer layers and the average $\delta_m$ can decrease because of increased Coulomb repulsion or correlation effects in the $(Cu)_{13}$ cluster as the number of layers in the cluster increases.

We have found that Eq. (5) correctly predicts the $T_c^m$ for several high-$n$ compounds. Using the $q_m$ and $\delta_m$ for LSCO and NCCO we have calculated the $T_c^m$ for the $n=\infty$ versions of these compounds $(Sr_{1-x}Ca_x)_{1-y}CuO_2$ (Ref. 17) and $Sr_{1-x}Nd_xCuO_2$,[18] as 96 and 45 K, respectively, in good agreement with the experimental values of 110 and 40 K, respectively. Similarly, for the 1-Tl homologous series ($a=3.9$, $c_n=9.7+(n-1)3.2$ Å), using the $q_m$ and $\delta_m$ for the $n=1$ member of the series $\{T_c^m(n=1)=50 \text{ K}\}$, we obtain $T_c^m=86$ ($n=2$), 110 ($n=3$), and 126 K ($n=4$) while experiment gives 80 ($n=2$), 110 ($n=3$), and 122 K ($n=4$).[19] The excellent agreement with experiment indicates that in the 1-Tl series $q_m$ and $\delta_m$ do not change appreciably with increasing $n$. The agreement is not quite as good for the 2-Tl, 2-Bi, and 1-Hg series, indicating that in these compounds, $q_m$ and/or $\delta_m$ decrease with increasing $n$.

## IV. SUPERFLUID DENSITY

The two principal ingredients of superconductivity are the phase coherence and the pairing strength of the superconducting charge pair. The phase coherence is related to the superfluid carrier density $n_s$, and the pairing strength is related to the superconducting energy gap $\Delta$. We discuss the implications of the $(Cu)_{13}$-BEC model on the superfluid density in this section and on the pairing strength in the next section.

In the model, the superfluid density is directly proportional to the dopant charge $q$. This is consistent with the concept that the cuprates are doped Mott insulators. The superfluid density $n_s$ is given by $n_s = 13nqB(T/T_c)/(13a^2c)$, where $B(T/T_c)$ is a Bose condensation factor. Although the cuprates are anisotropic with charge transport primarily within the CuO$_2$ layers, we use the homogeneous Bose condensation factor, $B(T/T_c) = [1-(T/T_c)^{3/2}]$,[11] since we do have conductivity along the $c$ axis in the normal state and tunneling interactions between layers in the superconducting state.

The superfluid density in the cuprates has been investigated by measurements of the penetration depth $\lambda_p$ by microwave and muon spin-relaxation ($\mu$SR) techniques. In particular, the temperature dependence of $n_s(T)/n_s(0)$ can be obtained from these two methods by measurements of $\lambda_p^2(0)/\lambda_p^2(T)$. For a given cuprate, $n_s$ varies as $13qB(T/T_c) = n_{sc}$, the number of carriers in the superconducting state in one layer of a $(Cu)_{13}$ cluster. As long as $q \leq \frac{2}{13}$, the underdoped and optimally doped regions, the temperature dependence will be given by $B(T/T_c)$. For $q > \frac{2}{13}$, the overdoped region, we need to keep in mind that $n_{sc}$ must be $\leq 2$. In Fig. 6 we show data for $\lambda_p^2(0)/\lambda_p^2(T)$, which is proportional to $n_s(T)/n_s(0)$, obtained by microwave and $\mu$SR measurements on optimally doped Y123 and Bi2212 samples,[20–22] together with the Bose condensation factor $B(T/T_c)$. The agreement with the Bose condensation factor is quite good up to $T/T_c$ of $\approx 0.4$. The variance at higher $T$ may be the result of the anisotropy of the cuprates, or the effects of fluctuations in the phase coherence that are particularly large in the cuprates because of the small value of the coherence length. Although the reasonable fit of $\lambda_p^2(0)/\lambda_p^2(T)$ to the Bose-Einstein condensation factor $B(T/T_c)$ does not in itself substantiate the specific $(Cu)_{13}$-BEC model herein proposed, it does provide some support for the role of a Bose-Einstein mechanism in the cuprates.

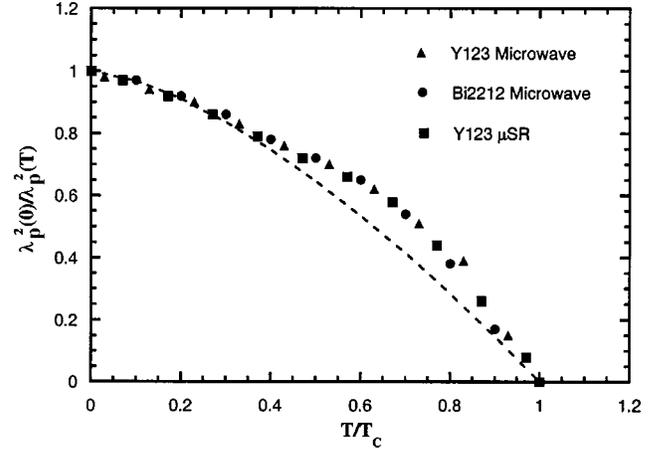

FIG. 6. Variation of superfluid density fraction $n_s(T)/n_s(0)$ with temperature. Experimental ratios of penetration depths $\lambda_p^2(0)/\lambda_p^2(T)$, which are proportional to $n_s(T)/n_s(0)$, for optimally doped Y123 and Bi2212 from microwave and $\mu$SR measurements are from Refs. 20–22. The Bose condensation factor $B(T/T_c) = [1-(T/T_c)^{3/2}]$ is represented by a dashed line.

The superconducting peak (SCP) seen in angle-resolved photoemission spectroscopy (ARPES) has been associated with the superconducting density as well.[23–25] In Fig. 7 we show the relative change in the spectral weight of the SCP feature, which is assumed to be proportional to the superfluid density, as a function of $q$ for Bi2212 samples at a temperature of 14 K.[24] We also plot the predicted change in $n_{sc} = 13qB(T/T_c)$, the number of superconducting carriers in a CuO$_2$ layer of the (Cu)$_{13}$ cluster. The theoretical curve increases rapidly from the threshold $q_0$ value of $\approx 0.06$ and then increases approximately linearly to the maximum value of 2 at $q \approx 0.16$. It then remains constant at 2 until $q \approx 0.25$, after which it drops rapidly to zero at $q \approx 0.26$, where $T_c$ drops below the experimental temperature of 14 K. Note that the theoretical curve in Fig. 7 is from the (Cu)$_{13}$-BEC model and thus the $q$ scale is related to the model $T_c$ vs $q$ curve. The original ARPES data is presented with a $q$ scale derived from the Presland empirical formula.[16]

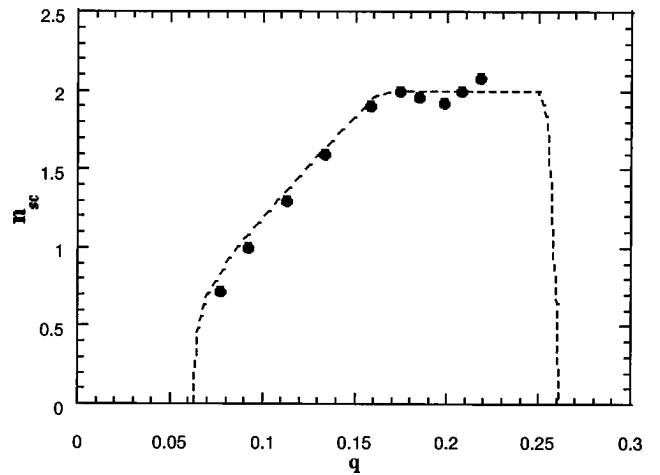

FIG. 7. The theoretical number of carriers in the superconducting state per layer in the $(Cu)_{13}$ cluster $n_{sc}$ at 14 K for Bi2212 (dashed line) vs the dopant charge per Cu $q$. The data is the relative spectral weight of the superconducting peak (SCP) in the ARPES spectrum at 14 K from Ref. 24. The data has been shifted to correspond to the appropriate model $q$ values for the different sample $T_c$'s.

We can see from Fig. 5 that, for the same value of $T_c$, there is in general a shift of between 0.005 and 0.01 in $q$ between the model curve and the empirical curve, with the model $q$ being lower than the empirical $q$ for $q > 0.07$. The ARPES data in Fig. 7 has therefore been shifted so as to place both theory and data on the same model $q$ scale. As seen in Fig. 7, the agreement between theory and experiment is very good. A further test of the (Cu)$_{13}$-BEC model would be to study the drop in $n_{sc}$ for $q > 0.25$, which by our model represents overdoped Bi2212 with $T_c/T_c^m < 0.33$.

The (Cu)$_{13}$-BEC model indicates that studies of $n_s(q)$ by $\mu$SR or microwave measurements will be complicated by the presence of normal carriers at $T=0$ in the overdoped region. For example, the muon spin-relaxation rate $\sigma$ will decrease when normal carriers are present,[26] and thus the rate should not simply vary with $n_s(0)$ but rather with a weighted superfluid density, $n'_s(0)$, where $n'_s(0) = n_s(0)\{n_s(0)/n(0)\}$ with $n(0)$ being the total density of all carriers present at $T=0$. Keeping in mind that $n_s$ is proportional to $n_{sc}$, we plot, in Fig. 8(a), $n_{sc}(0)$ and $n'_{sc}(0)$, which are given by $13qB$ and $n_{sc}(0)\{13qB/13q\}$, respectively, for a generic cuprate with a threshold $q_0$ of $\approx 0.06$. At $T=0$ the two quantities are identical linear functions of $q$ between 0.06 and 0.154 ($\frac{2}{13}$), the underdoped and optimally doped regions. In the overdoped region $q > \frac{2}{13}$, $n_{sc}(0)$ remains constant at 2, while $n'_{sc}(0)$ decreases. This linear increase followed by a decrease in the overdoped region has been observed in the $\mu$SR rate results on Y123,[25] indicating that the $\mu$SR signal is indeed affected by the presence of the normal carriers. However, the peak in Y123 is observed at an empirical $q \approx 0.19$, which translates to a model $q \approx 0.184$, still a fairly long way from $q = 0.154$. This, however, might be the result of the $c$-axis superconductivity that is present in Y123. A good test would be a systematic $\mu$SR vs $q$ study on Bi2212.

Let us now examine the well-known Uemura relationship. Uemura found that for a given cuprate there is a linear relationship between $T_c$ and $\sigma$, and thus between $T_c$ and $n_s/m^*$ ($m^*$ is the effective mass) in the underdoped region, followed by a saturation in the optimally doped region.[9] In the overdoped region there appears to be a restoration of the linear relationship, but now $\sigma$ decreases with increasing doping.[26] Since $m^*$ does not appear to change with doping,[26] this decrease in $\sigma$ appears to be the result of an apparent decrease in $n_s$. In Fig. 8(b) we plot the ratio of $T_c/n_{sc}(0)$ and $T_c/n'_{sc}(0)$ for the generic cuprate with a $T_c^m$ of 100 K. We see that $T_c/n_{sc}(0)$ is fairly constant in much of the underdoped region, indicating a linear relationship between $T_c$ and $n_{sc}$ and therefore between $T_c$ and $n_s/m^*$ (or $\sigma$). As $q$ approaches the optimally doped region, the ratio $T_c/n_{sc}$ begins to decrease, indicating a saturation. In the overdoped region, while $T_c/n_{sc}$ continues to decrease, $T_c/n'_{sc}$ remains fairly constant until $q \approx 0.22$. Since, as we discussed above, the spin-relaxation rate $\sigma$ should vary with the weighted superfluid density $n'_{sc}$, rather than $n_{sc}$, in the overdoped region, the linear relationship between $T_c$ and $\sigma$ is restored but $\sigma$ now decreases with increasing $q$. These results appear to be fully consistent with the findings of Uemura.

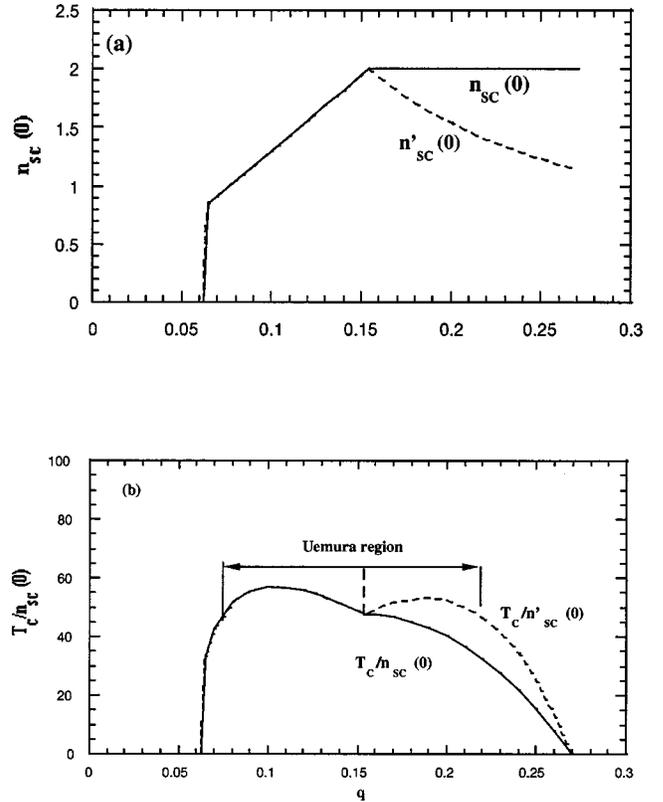

FIG. 8. (a) The theoretical number of carriers in the superconducting state per layer in the (Cu)$_{13}$ cluster at 0 K, $n_{sc}(0)$, for a generic cuprate (solid line) vs the dopant charge per Cu, $q$. The weighted number of superconducting carriers $n'_{sc} = n_{sc}(0) \times \{n_{sc}(0)/n(0)\}$, where $n(0)$ represents the total number of carriers in the layer per cluster at 0 K, is designated by a dashed line. (b) The calculated ratio $T_c/n_{sc}(0)$ (solid line) and $T_c/n'_{sc}(0)$ (dashed line) for a generic cuprate with a $T_c^m = 100$ K. Regions of approximately constant ratio indicate regions where there is an approximate linear relationship between $T_c$ and $n_{sc}$ or $n'_{sc}$ and thus between $T_c$ and the $\mu$SR rate $\sigma$.

The (Cu)$_{13}$-BEC model thus appears to account for the major characteristics of the superfluid density in the cuprates. In particular, it agrees with the temperature and doping dependencies observed in microwave, $\mu$SR, and ARPES stud-ies. The experimental results, in turn, are consistent with the principle of a $(Cu)_{13}$ cluster with one singlet superconducting state, and thus a maximum of two superconducting carriers per layer, and with a temperature dependence that is consistent with a Bose condensation factor. In addition, the model appears able to reproduce the Uemura $T_c$ vs $\sigma$ results for the cuprates.

## V. PAIRING STRENGTHS

### A. Normal state

In BCS superconductivity there are no preformed pairs and the bosonic Cooper pairs do not exist for $T > T_c$. Instead the formation of Cooper pairs and the condensation to the superconducting state occur simultaneously at $T = T_c$. In the $(Cu)_{13}$-BEC model we have bosonic preformed pairs at $T > T_c$ with a density that is a Boltzmann function of temperature in accordance with the occupation probability of the $(\psi_0)^2$ ground state of the $(Cu)_{13}$ cluster orbital energy manifold. The energy manifold, depicted in Fig. 3, has the Fermi level $E_F$ set at the center nonbonding orbital. The single-particle energy of the $(\psi_0)^2$ singlet state is at $-6\delta$. The other $n_n$ normal states of the $(Cu)_{13}$ manifold, $\psi_j$ ($j=1,2...12$), are separated by an energy $\delta$ from each other, and thus are at energies $-(6\delta-j)$. In the normal state, the number of charges in state $\psi_0$ per layer in a single $(Cu)_{13}$ cluster is given by $n_c(\psi_0) = 13qP(\delta,T)$ and the number in state $\psi_j$ by $n_c(\psi_j) = n_c(\psi_0)e^{-j\delta/kT}$. An anomalous characteristic of the normal state of the superconducting cuprates is the presence of a pseudogap. Experimentally the pseudogap is defined as the distance from $E_F$ to the leading-edge midpoint (LEM) of the ARPES spectrum when $T > T_c$. Since the $(\psi_0)^2$ singlet state is at energy $-6\delta$ relative to the Fermi energy, this then represents a pseudogap at $T > T_c$ in the density of states near $E_F$. Since the various states $\psi_j$ within the pseudogap are thermally populated, the pseudogap, as measured by the LEM, should decrease with increasing $T$, in agreement with observation.

Since the $(\psi_0)^2$ singlet state at $T > T_c$ is not a condensed coherent state, the pairing strength of the preformed pair is simply the energy $\delta$ required to promote one of the charges in the $(\psi_0)^2$ state into the next higher-energy state. Thus preformed pairs are weakly bonded pairs with a pairing energy in the 1–4-meV range and are not hard-core bosons. It is interesting to note that the pairing strengths of the normal-state preformed pairs in the cuprates are comparable to the pairing strengths of the superconducting Cooper pairs in the BCS superconductors. The $(Cu)_{13}$-BEC model may thus provide a basis for understanding the unusual normal properties of the superconducting cuprates. In the normal state the singlet charge pairs exist as preformed pairs with wave functions that extend $\approx 20$ Å, and that are confined primarily to the $CuO_2$ planes. The pairs have low binding energies, are weakly interacting, and have a density that is temperature dependent. The low binding energy of the charge pairs indicates that the preformed pairs will fluctuate, i.e., readily break up and reform under phonon interactions. The unusual normal properties of the superconducting cuprates, including their non-Fermi-liquid behavior, may be the result of the presence of these fluctuating preformed pairs. As discussed earlier, the theory further predicts that when $\delta$ goes to zero in the overdoped region, these preformed pairs will disappear and the cuprates will then behave as more conventional Fermi-liquid metals.

### B. Superconducting state

We have assumed that only the singlet $(\psi_0)^2$ states form bosonic pairs that can undergo Bose-Einstein condensation into coherent pairs. Given that the correlation length is only $\approx 2.75a$, there is essentially only one coherent pair within the correlation length, although there is some overlap between pairs. The superconducting pair thus acts as a single coherent quasiparticle $\psi_s$, located at $E \approx -12\delta$, as shown in Fig. 3 (actually this energy might be somewhat less than $12\delta$ because of electron correlation or Coulomb repulsion effects). It is this coherent quasiparticle that gives rise to the superconducting peak (SCP) in the ARPES spectra, and which, according to the model, will have a superconducting gap $\Delta \approx 12\delta$. The LEM is derived from the spectral envelope of all states near the Fermi edge. At $T=0$, this envelope is set primarily by the SCP. At higher temperatures, the envelope includes the normal states as well. The superconducting gap $\Delta$, the LEM, and the pseudogap all arise from a common origin, the preformed pair, and all $\sim \delta$. Therefore all three will have the same $d_{x^2-y^2}$ symmetry, in agreement with experiment.

Since $\delta$ is not a function of $T$ but is a linearly decreasing function of $q$, we expect that $\Delta$ will have the exact same dependencies. This is indeed in agreement with ARPES data that show that the SCP energy $\Delta$ does not change with $T$ but does decrease linearly with $q$.[24] We list in Table II the calculated values for $\Delta$ at optimal doping, $\Delta_m = 12\delta_m$, for the representative cuprates along with measured or estimated values.[25] The agreement is quite good. Besides providing reasonable agreement on the absolute magnitudes, the $(Cu)_{13}$-BEC model also accounts for the observed linear decrease in the superconducting gap, the leading-edge midpoint, and the pseudogap with increased doping since all three pairing strengths ($\Delta$, LEM, and pseudogap) are proportional to $\delta$ and $\delta$ itself decreases linearly with $q$.

Experiments on the dependence of the pairing strengths on temperature indicate that the superconducting gap $\Delta$ and the LEM depend on $T$ differently in underdoped and optimally doped samples than in overdoped samples.[27–30] In underdoped and optimally doped samples, $\Delta$ is essentially independent of $T$ (up to $T_c$), while the LEM decreases slowly with $T$ between 0 and $T_c$ and than more rapidly for $T > T_c$. However, for these materials, the LEM does not approach 0 until $T \gg T_c$, thus leading to the presence of a $T > T_c$ pseudogap. In overdoped samples, $\Delta$ shows some small apparent decrease with $T$ as $T \to T_c$, and the LEM in these samples decreases rapidly with $T$ for $T > 0$, reaching zero near $T_c$. Thus overdoped samples do not exhibit a $T > T_c$ pseudogap.

To understand the temperature dependencies of the pairing strengths as predicted by the $(Cu)_{13}$-BEC model we il-lustrate in Fig. 9 the predicted evolution of the electronic states with temperature. In Fig. 9(a) we consider an underdoped sample, while Fig. 9(b) deals with an overdoped sample. In the underdoped sample at $T/T_c = 0$, all of the spectral weight resides in the SCP and there are no normal carriers. At $T/T_c = 0.75$, the SCP has decreased and the lowest-lying normal states begin to be populated. For $T > T_c$, the SCP disappears but still only a few of the normal states are populated because of the relatively large value of $\delta$. Figure 9(a) shows that the superconducting gap $\Delta$ will be independent of $T$ and exist up to $T_c$. On the other hand, the LEM will show some decrease with $T$ between 0 and $T_c$, and then a more rapid decrease for $T > T_c$. However, for the underdoped sample it is clear that the LEM will not go to zero until $T \gg T_c$. The value of the LEM for $T > T_c$ defines the pseudogap, and as we can see in Fig. 9(a), underdoped and optimally doped samples will exhibit a $T > T_c$ pseudogap which will decrease with increasing temperature.

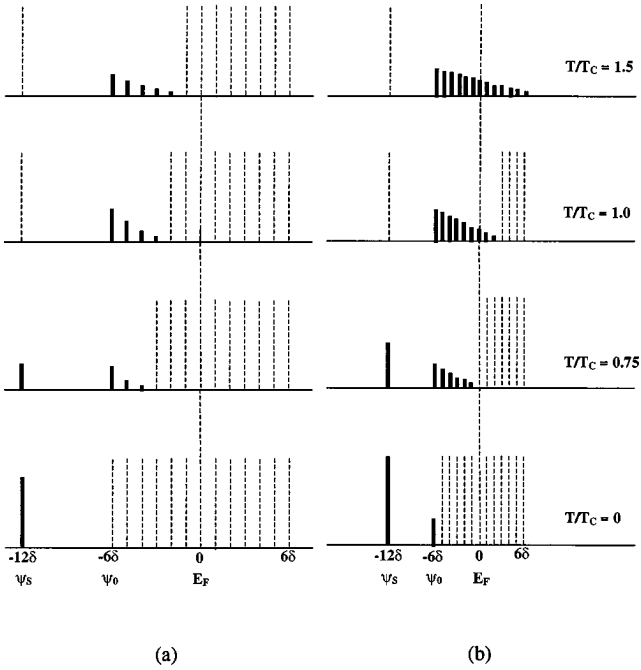

FIG. 9. (a) Illustrative depiction of the theoretical temperature evolution of the electronic states in an underdoped cuprate. The superconducting gap $\Delta$ is at $-12\delta$, the position of the superconducting coherent quasiparticle peak $\psi_s$. The leading-edge midpoint (LEM) decreases with $T$ as the normal states become thermally populated, but does not go to zero until $T \gg T_c$, thus producing a $T > T_c$ pseudogap. (b) The same for an overdoped sample. The overdoped sample has a bigger superconducting peak at $-12\delta$, but $\delta$ is much smaller than for the underdoped sample. The LEM decreases much more rapidly and reaches zero at $T \approx T_c$, thus producing no $T > T_c$ pseudogap.

Overdoped samples show somewhat different behavior as illustrated in Fig. 9(b). First, since $\delta$ is considerably smaller for overdoped samples, $\Delta$ is proportionately smaller as well. Also, the spectral weight of the SCP is greater than in the underdoped sample since $q$ is larger. In addition, there are now $13(q - \frac{2}{13})$ normal carriers per cluster layer present in the $\psi_0$ state at $T = 0$. As $T$ is increased, the normal states become populated at a faster rate than in the underdoped case because of the decreased value of $\delta$ and the presence of normal carriers in $\psi_0$ at $T = 0$. The closer proximity of the $\psi_s$ and $\psi_0$ states and the presence of a considerable number of normal carriers at low temperatures in the overdoped samples can lead to some apparent decrease in the measured position of $\Delta$ as $T \rightarrow T_c$. More importantly, the LEM decreases much more rapidly and reaches zero in this example near $T_c$. This accounts for the apparent absence of a $T > T_c$ pseudogap in overdoped samples.

While detailed comparisons to actual ARPES spectra will require incorporation into the simulation of the valence electron background, the various linewidths, and other spectral features, such as the high-energy pseudogap feature, the $(Cu)_{13}$-BEC model appears to be consistent with the key results related to the temperature dependence of the superconducting peak, superconducting gap, leading-edge midpoint and pseudogap obtained from ARPES spectra. It is particularly gratifying that the same magnitudes and dopant dependence of the preformed pair energy $\delta$ that we find best agree with the key thermodynamic property, the $T_c$ vs $q$ curve, also agree with the key electronic properties of the superconducting cuprates.

## VI. CONCLUSIONS

Let us summarize the various characteristics of high-temperature superconductors for which the $(Cu)_{13}$-BEC model is able to account. The model provides a basis for understanding the normal-state properties in these materials through the presence of fluctuating preformed pairs. The dominance of the $CuO_2$ planes and thus the anisotropy of the cuprates are fundamental aspects of the theory. The model provides a superconducting pair with the right wave-function symmetry $d_{x^2-y^2}$ and the right correlation length $\approx 10$ Å. It accounts for the unusually high $T_c$'s of the cuprates through a BEC mechanism in systems that have relatively high superfluid densities and pairing strengths. The model correctly predicts that the superconductivity threshold doping levels should be at $\approx 0.05$ charges per in-plane Cu. It reproduces the bell-shaped $T_c$ doping curves for the cuprates and accounts for them through the effects of an increasing and then saturating superfluid density coupled with a decreasing pairing strength. The model provides natural explanations for why the maximum $T_c$'s for the superconducting cuprates tend to occur near dopant concentrations of 0.15–0.16 charges per in-plane Cu, and why the superconductivity disappears, and a Fermi-liquid state emerges, in the overdoped region. It also explains why maximum $T_c$'s first increase and then saturate, or decrease, as the number of $CuO_2$ planes in a unit cell increases, and provides good agreement with data for several high-$n$ materials. It is also able to reproduce the Uemura results on the $T_c$ dependence on the $\mu$SR rate for the cuprates. The model is able to account for the magnitudes and symmetries of the superconducting gaps, leading-edge midpoints, and the pseudogaps. Perhaps, most importantly, the model is also able to account for many of the key experimental ARPES, $\mu$SR, and microwave results on the doping and temperature dependencies of both the superfluid density and the pairing strengths (superconducting gap, leading-edge midpoint, and pseudogap) in the cuprates.

The $(Cu)_{13}$-BEC model thus appears to be quite promising, has a number of very attractive features, and is able to account quantitatively for many of the thermodynamic and electronic characteristics of the superconducting cuprates. Although this model was specifically developed for the cuprates, the general concepts of this model can be extended to other materials, including nonlayered systems. It may well be possible that other compounds can form hybridized preformed pairs that are spatially bound, have relatively low binding energies, and are weakly interacting and thus able to undergo Bose-Einstein condensation. If the densities of these preformed pairs is high enough and the pairing energies not too low, then these compounds may exhibit transition temperatures comparable to or higher than the cuprates.

## ACKNOWLEDGMENTS

I thank T. Geballe, R. Birgeneau, R. Laughlin, M. Beasley, and A. Kapitulnik for helpful discussions.